\documentclass[aps,prd,preprint,12pt]{revtex4}
\usepackage{graphicx,amssymb,amsfonts}
\newcommand{\be}{\begin{equation}}
\newcommand{\ee}{\end{equation}}
\newcommand{\cM}{{{\cal{M}}}}
\newcommand{\tw}{\tan\theta_W}

\newcommand{\bea}{\begin{eqnarray}}
\newcommand{\eea}{\end{eqnarray}}
\begin{document}
\begin{center}
\vspace{3cm}
\end{center}
%\begin{flushright}
%IZTECH-P-08-04
%\end{flushright}
\title{\Large \bf  Electric Dipole Moments in U(1)$^{\prime}$
Models\\}
\author{ Alper Hayreter$^{a,c}$, Asl{\i} Sabanc{\i}$^{a}$, Levent Solmaz$^{b}$, Saime Solmaz$^b$}
\affiliation{$^a$ Department of Physics, Izmir Institute of
Technology, IZTECH, Turkey, TR35430}

\affiliation{$^b$ Department of Physics, Bal{\i}kesir University,
Bal{\i}kesir, Turkey, TR10145}

\affiliation{$^c$ Department of Physics, Concordia University, 7141 Sherbrooke West,
Montreal, Quebec, Canada, H4B 1R6}

\date{\today}

\begin{abstract}
\medskip
We study electric dipole moments (EDM) of electron and proton in
E(6)--inspired supersymmetric models with an extra U(1)
invariance. Compared to the Minimal Supersymmetric Standard Model
(MSSM), in addition to offering a natural solution to the $\mu$
problem and predicting a larger mass for the lightest Higgs boson,
these models are found to yield suppressed EDMs.
\end{abstract}

\maketitle

\section{Introduction}
While solving the quadratic divergence of radiative corrections to the Higgs boson mass,
 the supersymmetrization of the Standard Model with minimal matter content  brings a $\mu$ parameter
 with a completely unknown scale. On the other hand,
 extending the gauge structure $SU(3)_C \times SU(2)_L \times U(1)_Y$
 of the Minimal Supersymmetric Model by a new $U(1)$ Abelian group
  provides an effective $\mu$ term related with the VEV of some extra singlet scalar field;
   thus a scale ($\sim TeV$) can be dynamically generated for the $\mu$ parameter.
The supersymmetric $U(1)^{\prime}$ models have been intensely
studied in the literature. While such models can be motivated by
low-energy arguments like $\mu$ problem \cite{muprob} of the MSSM
they also arise at low-energies as remnants of GUTs such as
$SO(10)$ and $E(6)$
\cite{Robinett:1981yz,Robinett:1982tq,Langacker:1984dc}. These
models necessarily involve an extra neutral vector boson
 \cite{Cvetic:1995rj,Cvetic:1996mf} whose absence/presence to be
 established at the LHC.

The particle spectrum of $U(1)^{\prime}$ models involve  bosonic
fields $Z_{\mu}^{\prime}$ and $S$ as well as their superpartners
$\widetilde{Z}^{\prime}$ and $\widetilde{S}$ in addition to those
in the MSSM. Therefore, such models can be tested in various
observables ranging from electroweak precision observables to
$Z_{\mu}^{\prime}$ effects at the LHC. As a matter of fact,
analysis of Higgs sector along with CP violation potential
 \cite{Demir:2005kg} as well as structure of EDMs \cite{Suematsu:1998wm}
 suggest several interesting signatures also at collider experiments
\cite{King:2005jy}. One of the most important spots of these
models is that the lower bound of the lightest Higgs boson mass
($m_h \geq$114 GeV) can be satisfied already at the tree level,
and radiative corrections (dominantly the top--stop mass
splitting) is not needed to be as large as in the MSSM. This
feature can have important implications also for the little
hierarchy problem \cite{Demir:2005ti}.

In this work we will study EDMs of electron and neutron in
$U(1)^{\prime}$ models stemming from $E(6)$ GUT. Our main interest
is to look at the reaction of EDMs to gauge extensions in
comparison to the MSSM. The  paper is organized as follows.   In
the  next section we introduce the models. Section III is devoted
to EDM predictions and their numerical analysis. In Section IV we
conclude.

\section{The $U(1)^{\prime}$ Models}
The model is characterized by the gauge structure \bea SU(3)_C
\times SU(2)_L \times U(1)_Y \times U(1)_{Y^{\prime}} \eea where
$g_3$, $g_2$, $g_Y$ and $g_{Y^{\prime}}$ are gauge coupling
constants respectively. Here the extra $U(1)$ symmetry can be a
light (broken at a $\rm{TeV}$) linear combination of a number of
U(1) symmetries (in effective string models there are several U(1)
factors whose at least one combination can survive down to the
$\rm {TeV}$ scale). There are a number of $U(1)^{\prime}$ models
studied in literature, all of them offer a dynamical
 solution to the $\mu$ problem of the MSSM via spontaneous breaking of extra $U(1)$ Abelian
 factor at the $\rm{TeV}$ scale depending on the model, and many of them respecting gauge couplings
 unification predicts extra fields in order to sort out gauge and
 gravitational anomalies from the theory. These models typically
 arise from SUSY GUTs and strings. From $E(6)$ GUT, for example,
two extra $U(1)$ symmetries appear in the breaking
  $E6 \rightarrow SO(10) \times U(1)_{\psi}$ followed by $SO(10) \rightarrow SU(5) \times U(1)_{\chi}$
   where $U(1)_{Y^{\prime}}$ is a linear combination of $\psi$ and $\chi$
   symmetries:
\bea U(1)_{Y^{\prime}} = \cos{\theta_{E6}}\,U(1)_{\chi} -
\sin{\theta_{E6}}\,U(1)_{\psi} \eea which, supposedly, is broken
spontaneously at a ${\rm TeV}$. There arises, in fact, a continuum
of $U(1)^{\prime}$ models depending on the value of mixing angle
$\theta_{E_6}$. However, for convenience and traditional reasons,
one can pick up specific values of $\theta_{E_6}$ to form a set of
models serving a testing ground. We thus collected some well-known
models in Table \ref{tab1} with the relevant normalization factors
and a common gauge coupling constant \bea g_{Y^{\prime}} =
\sqrt{\frac{5}{3}} g_2 \tan{\theta_W} \eea

\begin{table}[h]
\centering
\begin{tabular}{c|c|c|c|c|c}
&$2\sqrt{15}~Q_{\eta}$ & $~~2~Q_I~~$& $2\sqrt{6}~Q_{\psi}$ &
$2\sqrt{10}~Q_N$ & $2\sqrt{15}~Q_S$
\\ \hline
  $u_L , d_L$    & -2    & ~~0  & ~~1 & ~~1 & -1/2 \\
  $u_R$          & ~~2   & ~~0  & -1  & -1  & ~~1/2  \\
  $d_R$          & -1    & ~~1  & -1  & -2  & -4  \\
  $e_L$          & ~~1   & -1   & ~~1 & ~~2 & ~~4  \\
  $e_R$          & ~~2   & ~~0  & -1  & -1  & ~~1/2  \\
  $H_u$          & ~~4   & ~~0  & -2  & -2  & ~~1  \\
  $H_d$          & ~~1   & ~~1  & -2  & -3  & -7/2  \\
  $S$            & -5    & -1   & ~~4 & ~~5 & ~~5/2   \\
\end{tabular}
\caption{Gauge quantum numbers of several $U(1)^{\prime}$ models
\cite{Langacker:2008yv}} \label{tab1}
\end{table}

In theories involving more than one $U(1)$ factor the kinetic terms
can mix since for such symmetries the field strength tensor itself
is invariant. In $U(1)^{\prime}$ model, involving hypercharge
$U(1)_Y$ and $U(1)_{Y^{\prime}}$, the gauge part of the Lagrangian
takes the form \bea -{\cal L}_{gauge} = \frac{1}{4} F_{Y}^{\mu
\nu} F_{Y \mu \nu} +\frac{1}{4} F_{Y^{\prime}}^{\mu \nu}
F_{Y^{\prime} \mu \nu} +\frac{\sin{\chi}}{2} F_{Y}^{\mu \nu}
F_{Y^{\prime} \mu \nu} \eea where $F_{\mu \nu} = \partial_{\mu}
Z_{\nu} -
\partial_{\nu} Z_{\mu}$ is the field strength tensor of the
corresponding $U(1)$ symmetry. Kinetic part of Lagrangian can be
brought into canonical form by a non-unitary transformation \bea
\left(
\begin{array}{c} \hat{W}_{Y} \\  \hat{W}_{Y^{\prime}}
\end{array} \right) =
\left( \begin{array}{c c}
1 & -\tan{\chi} \\
0 & 1/\cos{\chi}
\end{array} \right)
\left( \begin{array}{c}
\hat{W}_{B} \\ \hat{W}_{B^{\prime}}
\end{array} \right)
\eea
where $\hat{W}_{Y}$ and $\hat{W}_{Y^{\prime}}$ are the chiral superfields
associated with the two $U(1)$ gauge symmetries. This transformation also acts
 on the gauge boson and gaugino components of the chiral superfields in the same form.
 The $U(1)_Y \times U(1)_{Y^\prime}$ part of covariant derivative in the case of no kinetic mixing is given by
\bea D_{\mu} = \partial_{\mu} + i g_Y Y B_{\mu} + i g_{Y^{\prime}}
Q_{Y^{\prime}} B_{\mu}^{\prime} \eea however, with the presence of
kinetic mixing this covariant derivative is changed to \bea D_{\mu}
= \partial _{\mu} + i g_Y Y B_{\mu} + i\left( -g_Y Y \tan{\chi} +
\frac{g_{Y^{\prime}}}{\cos{\chi}} Q_{Y^{\prime}} \right)
B_{\mu}^{\prime} \eea where $g_{Y^{\prime}}$ is gauge coupling
constant and $Q_{Y^{\prime}}$ is fermion charges of
$U(1)_{Y^{\prime}}$ symmetry. With a linear transformation of
charges the covariant derivative takes the form \cite{Choi:2006fz}
\bea D_{\mu} = \partial_{\mu} + i g_Y Y B_{\mu} + i g_{Y^{\prime}}
Q_{Y^{\prime}}^{\prime} B_{\mu}^{\prime} \eea in which the effective
$U(1)_{Y^{\prime}}$ charges are shifted from its original value
$Q_{Y^{\prime}}$ to \bea Q_{Y^{\prime}}^{\prime} =
\frac{Q_{Y^{\prime}}}{\cos{\chi}} - \frac{g_Y}{g_{Y^{\prime}}} Y
\tan{\chi} \eea

For the proper treatment of the models the most general
superpotential should be considered \cite{King:2005jy}, but for
simplicity we parametrized $U(1)^{\prime}$ models by the following
superpotential \bea \widehat{W} =  h_u \widehat{Q} \cdot
\widehat{H}_u  \widehat{U^c} + h_d \widehat{Q} \cdot \widehat{H}_d
\widehat{D^c} + h_e \widehat{L} \cdot \widehat{H}_d \widehat{E^c}
+ h_S \widehat{S} \widehat{H}_u \cdot \widehat{H}_d \eea where we
discarded additional field (assuming that they are relatively
heavy compared to this very spectrum) that are necessary for the
unification of gauge couplings. Our conventions are such that, for
instance $\widehat{Q} \cdot \widehat{H}_u \equiv \widehat{Q}^{T}
\left(i \sigma_2\right) \widehat{H}_u = \epsilon_{i j}
\widehat{Q}^i \widehat{H}^j_u$ with
$\epsilon_{12}=-\epsilon_{21}=1$. The right-handed fermions are
contained in the chiral superfields $\widehat{U}$, $\widehat{D}$,
$\widehat{E}$ via their charge-conjugates $e.g.$ $\widehat{U} =
\left(\widetilde{u_R}^{\star}, \left(u_R\right)^C\right)$. What a
$U(1)^{\prime}$ model does is basically to allow a dynamical
effective $\mu_{eff}=h_s \langle S \rangle$ related to the scale
of $U(1)^{\prime}$ breaking instead of an elementary $\mu$ term
which troubles supersymmetric Higgsino mass in the MSSM. Notice
that a  bare $\mu$ term cannot appear in the superpotential due to
$U(1)^{\prime}$ invariance.

At this point, it is useful to explicitly state the soft breaking terms, the most general holomorphic structures are
\bea \label{soft} \nonumber
-{\mathcal{L}}_{soft}&=&(\sum_i M_i \lambda_i\lambda_i-A_Sh_sSH_dH_u -A_u^{ij}h_u^{ij} U^c_jQ_iH_u\\ \nonumber
 &-& A_{d}^{ij}h_d^{ij} D^c_jQ_iH_d-A_e^{ij}h_e^{ij} E^c_jL_iH_d+h.c. )  \\ \nonumber
 &+& m_{H_u}^2|H_u|^2+m_{H_d}^2|H_d|^2+m_S^2|S|^2 \\ \nonumber
 &+&  m_{Q_{ij}}^2 \widetilde Q_i\widetilde Q_j^*+m_{U_{ij}}^2 \widetilde U^c_i\widetilde U^{c*}_j+m_{D_{ij}}^2 \widetilde D^c_i\widetilde D^{c*}_j+m_{L_{ij}}^2 \widetilde L_i\widetilde L_j^*\\
 &+&m_{E_{ij}}^2 \widetilde E^c_i\widetilde E^{c*}_j+h.c.
\eea
where the sfermion mass-squareds $m^2_{Q,...,E^c}$ and trilinear couplings
$A_{u,...,e}$ are $3\times3$ matrices in flavor space.
 All these soft masses will be taken here to be diagonal. In
 general, all gaugino masses, trilinear couplings and
 flavor-violating entries of the sfermion mass-squared matrices
 are source of CP violation. However, for
 simplicity and definiteness we will assume a basis in which
 entire  CP violating effects are confined into the gaugino mass $M_1$
(with  $M_1=M^{\prime}_1$), and the rest are all real
   (interested readers can chief to \cite{Demir:2007dt}).

These soft SUSY breaking parameters are generically nonuniversal
at low energies. We will not address the origin of these low
energy parameters as to how they follow via RG evolution from high
energy boundary conditions, instead we will perform a general scan
of the parameter space.

\section{Constraints and Implications for EDMs}
Due to the extra $U(1)$ symmetry, associated $Z^{\prime}$ boson
can be expected to weigh around the electroweak bosons, and can
exhibit significant mixing with the ordinary $Z$ boson. The LEP
data and other low-energy observables forbid $Z$--$Z^{\prime}$
mixing to exceed one per mill level. Indeed, precision
measurements have shown that $Z^{\prime}$ mass should not
  be less than $\sim 700$ GeV for any of the models under concern (excluding leptophobic $Z^{\prime}$'s).
   Indeed, mixing of the $Z$ and $Z^{\prime}$ puts important restrictions on the mass
   and the mixing angle of the extra boson and this can be studied from the following $Z-Z^{\prime}$ mixing matrix;
\bea
M_{Z-Z^{\prime}}^2=
\left( \begin{array}{c c}
M_Z^2 & \Delta^2 \\
\Delta^2 & M_{Z^{\prime}}^2 \\
\end{array}
 \right)
\eea
with $M_Z$ being the usual SM $Z$ mass in the absence of mixing and
\bea
M_Z^2 &=& \frac{1}{4} G^2 (|v_u|^2 + |v_d|^2) \nonumber \\
\Delta^2 &=& \frac{1}{2} G g_{Y^{\prime}} (Q^{\prime}_{H_u} |v_u|^2 - Q^{\prime}_{H_d} |v_d|^2) \\
M_{Z^{\prime}}^2 &=& g_{Y^{\prime}}^2 ({Q^{\prime}}^2_{H_u} |v_u|^2 + {Q^{\prime}}^2_{H_d} |v_d|^2 + {Q^{\prime}}^2_s |v_S|^2) \nonumber
\eea
where $G^2 = g_Y^2 + g_2^2$ and $g_{Y^{\prime}}$ is the gauge coupling constant of the extra $U(1)$.
The mixing matrix can be diagonalized by an orthogonal transformation;
\bea
\left( \begin{array}{c}
Z_1 \\ Z_2 \end{array} \right) =
\left( \begin{array}{c c}
\cos{\phi} & \sin{\phi} \\
-\sin{\phi} & \cos{\phi} \end{array} \right) \left( \begin{array}{c}
Z \\ Z^{\prime} \end{array} \right) \eea giving the mass eigenstates
$Z_{1,2}$ with masses $M_{Z_1,Z_2}$ where $\alpha$ is given by \bea
\tan{2\alpha} = \frac{2 \Delta^2}{M_Z^2 - M_{Z^{\prime}}^2} \eea In
the numerical analysis we considered
$\alpha< 3\times 10^{-3}$ and
confined $M_{Z^{\prime}}>700$ GeV. Notice that when $\Delta$
vanishes ($\tan{\beta} \sim
\sqrt{Q^{\prime}_{H_u}/Q^{\prime}_{H_d}}$) $Z_{1,2}$ can be
identified with the ordinary $Z$ and $Z^{\prime}$ bosons; since we
considered low $\tan{\beta}$ values, we will use the term
$Z^{\prime}$ for the heavy extra boson.

Besides this, the implication of the extra gauge boson can also be
seen in sfermion sector, that is sfermion mass matrix is modified due
 to the presence of $Z^{\prime}$ boson as;
\bea
\cM_{\tilde f}^{2}=
\left(
\begin{array}{cc}
\cM_{\tilde f_{LL}}^2                     &    \cM_{\tilde f_{LR}}^2\\
{\cM_{\tilde f_{LR}}^{2\star}}  &    \cM_{\tilde f_{RR}}^2
\end{array}
\right) \nonumber \\
\eea
\bea
\cM_{\tilde f_{LL}}^2 &=& \cM_{\tilde f_L}^2 + h_f^2 |H_f^0|^2 + \frac{1}{2} \left(Y_{f_L} g_Y^2 - T_{3f} g_2^2 \right) \left( |H_u^0|^2 - |H_d^0|^2 \right) \nonumber \\
&+& g_{Y^{\prime}}^2 Q_{f_L}^{\prime} \left( |H_u^0|^2 Q_{H_u}^{\prime} +
|H_d^0|^2 Q_{H_d}^{\prime} + |S|^2 Q_s^{\prime}   \right) \nonumber \\
\cM_{\tilde f_{LR}}^2 &=& h_f \left(A_f^{\star} {H_f^0}^{\star} + h_s S H_f^0 \right) \\
\cM_{\tilde f_{RR}}^2 &=& \cM_{\tilde f_R}^2 + h_f^2 |H_f^0|^2 + \frac{1}{2} \left( Y_{f_R} g_Y^2 \right) \left( |H_u^0|^2 - |H_d^0|^2 \right) \nonumber \\
&+& g_{Y^{\prime}}^2 Q_{f_R}^{\prime} \left( |H_u^0|^2 Q_{H_u}^{\prime} +
|H_d^0|^2 Q_{H_d}^{\prime} + |S|^2 Q_s^{\prime} \right) \nonumber
\eea
in terms of shifted charge assignments. Sfermion mass matrix is hermitian
 and can be diagonalized by the unitary transformation
\bea
D^{\dagger} {\cal M}_{\tilde f}^2 D = diag(m_{\tilde f_1}^2 , m_{\tilde f_2}^2)
\eea
where $D$ is the $L-R$ mixing matrix for sfermions and is parametrized as
\bea
 D =
 \left( \begin{array}{cc}
    \cos{\theta} & \sin{\theta}\, e^{-i\phi} \\
    \sin{\theta}\, e^{i\phi} & \cos{\theta} \\
  \end{array} \right)
\eea It is worth to note that sfermion mass eigenvalues in
$U(1)^{\prime}$ models will be different than in the MSSM due to
the contribution of extra gauge boson and kinetic mixing. In
general $D_{U(1)^\prime}\neq\,D_{MSSM}$ and the MSSM results can
be recovered by assuming no kinetic mixing ($\sin{\chi} = 0$) and
no charges under $U(1)^{\prime}$ at all.

\begin{figure}[ht!]
\begin{center}
\includegraphics[scale=0.6,height=35mm]{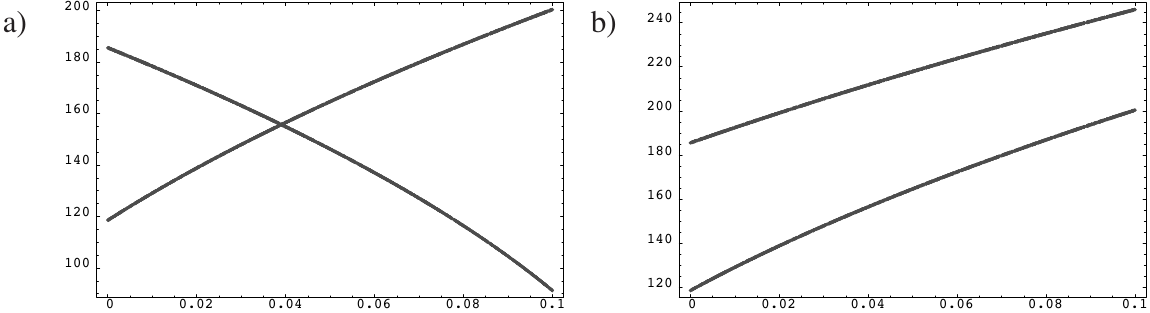}
\end{center}
\caption{Impact of selectron $U(1)^{\prime}$ charge Q$^\prime$ on
the selectron masses (In GeVs).} \label{fig1}
\end{figure}

But the existence of the $U(1)^{\prime}$ charges have profound
impact on the sfermion eigenvalues. To show this we present Fig.
\ref{fig1} in which selectron mass eigenvalues are plotted against
$U(1)^{\prime}$ charges for  two different cases. In panel a)
we assumed $Q^\prime_{eL}=-\,Q^\prime_{eR}=Q^\prime$
to be compared with panel b) in which
$Q^\prime_{eL}=Q^\prime_{eR}=Q^\prime$, with the following inputs:
$h_s=0.5$, $v_s=5$ TeV,  $Q^{\prime}_{H_u}=Q^{\prime}_{H_d}=-0.05$
 and the rest of the parameters are taken as in SPS1a$^\prime$ reference point
  \cite{AguilarSaavedra:2005pw},
and additionally we assumed  $A_s=A_t$. Notice that $Q^\prime=0$
corresponds to MSSM prediction. This figure illustrates the
difference between the MSSM and of the $U(1)^\prime$ sfermion mass
predictions, for the same input parameters. As should be inferred
from this figure, opposite values of $Q^\prime_{fL}$ and
$Q^\prime_{fR}$ can violate collider bounds for some of the
$U(1)^\prime$ models while this selection is current for the MSSM,
that will be important in the numerical analysis and we will
consider somewhat larger values of sfermion gauge eigenstates to
overcome this issue.

In $U(1)^{\prime}$ models compared to MSSM, there is an extra single
scalar state in Higgs sector, an additional pair of higgsino and
gaugino states are covered in neutralino sector and chargino sector
is kept structurally unaltered though it is different than the MSSM
due to the effective $\mu$ term. Now we will  deal with these
sectors.

\subsection{Higgs Sector}

The Higgs sector in $U(1)^{\prime}$ models compared to MSSM is
extended by a single scalar state S whose VEV breaks the
$U(1)^{\prime}$ symmetry and generates a dynamical $\mu_{eff}=h_S
\langle S \rangle$. For a detailed analysis of the Higgs sector
with CP violating phases we refer to \cite{Demir:2003ke} and
references therein. The tree level Higgs potential gets
contributions from F terms, D terms and soft supersymmetry
breaking terms:

\begin{equation}
\label{treepot} V_{tree}=V_F+V_D+V_{soft},
\end{equation}
in which
\begin{eqnarray}
\label{vf} V_F&=& |h_{s}|^2\left[ |H_u\cdot H_d|^2+ |S|^2 (
|H_u|^2+|H_d|^2)\right],
\end{eqnarray}
\begin{eqnarray}
\label{vd} V_D&=&\frac{G^2}{8}\left(|H_u|^2-|H_d|^2\right)^2+
   \frac{g_{2}^2}{2}(|H_u|^2|H_d|^2-|H_u \cdot H_d|^2)\nonumber \\
&+&\frac{g_{Y'}^2}{2}\left(Q^{\prime}_{H_u}|H_u|^2+Q^{\prime}_{H_d}|H_d|^2+
   Q^{\prime}_{S}|S|^2\right)^2,
\end{eqnarray}
\begin{equation}
V_{soft}=m_{u}^{2}|H_u|^2+
  m_{d}^{2}|H_d|^2+
  m_s^{2}|S|^2 +(A_sh_sSH_u\cdot H_d+h.c.).
\end{equation}
where $G^{2}=g_2^{2}+g_Y^{2}$ and $g_Y=\sqrt{3/5} g_1$, $g_1$ is the
GUT normalized hypercharge coupling.

At the minimum of the potential, the Higgs fields  can be expanded
as follows (see  \cite{Cvetic:1997ky} for a detailed discussion):
\begin{eqnarray}
\label{higgsexp} \langle H_u \rangle &=&
\frac{1}{\sqrt{2}}\left(\begin{array}{c} \sqrt{2} H_u^{+}\\ v_u +
\phi_u + i \varphi_u\end{array}\right)  \:,\:\:\:\: \langle H_d
\rangle = \frac{1}{\sqrt{2}}\left(\begin{array}{c} v_d +
\phi_d + i \varphi_d\\ \sqrt{2} H_d^{-}\end{array}\right)\nonumber\\
\langle S \rangle &=& \frac{1}{\sqrt{2}}  \left(v_s + \phi_s + i
\varphi_s\right),
\end{eqnarray}
in which $v^2\equiv v_u^2+v_d^2=(246\,{\rm GeV})^2$.  In the above
expressions, a phase shift $e^{i\theta}$  can be attached to $\langle
S \rangle$ which can be fixed by true vacuum conditions
considering loop effects (see \cite{Demir:2003ke} for details).
Here it suffices to state that the spectrum of physical Higgs
bosons consist of three neutral scalars (h, H, H$^\prime$), one CP
odd pseudoscalar (A) and a pair of charged Higgses H$^\pm$ in the
CP conserving case. In total, the spectrum differs from that of
the MSSM by one extra CP-even scalar.

Notice that, the composition,  mass and hence the couplings of the
lightest Higgs boson of $U(1)^{\prime}$ models can exhibit
significant differences from the MSSM, and this could be an
important source of signatures in the forthcoming experiments. It
is necessary to emphasize that these models can predict larger
values for $m_h$, which hopefully will be probed in near future at
the LHC. In the numerical analysis we considered $m_h>90$ GeV as
the lower limit. Besides this, as we will see, it is possible to
obtain larger values such as $m_h\sim 140$ GeV within some of
these E(6) based models.

\subsection{Neutralino Sector}
In $U(1)^{\prime}$ models the neutralino sector of the MSSM gets
enlarged by a pair of higgsino and gaugino states, namely $\tilde
S$ (which we call as `singlino') and ${\tilde B}^{\prime}$ (which
we call as bino-prime or zino-prime depending on the state under
concern). The mass matrix for the six neutralinos in the $({\tilde
B} , {\tilde W}^3 , {\tilde H}^0_d , {\tilde H}^0_u , {\tilde S} ,
{\tilde B}^{\prime})$ basis is given by
%% NEUTRALINO MASS MATRIX
\bea %\footnotesize
M_{\chi^0}=\left( \begin{array}{c c c c c c c c }
M_1 &   0 & -m_Z c_{\beta} s_{W}  & m_Z s_{\beta} s_{W}  & 0  & M_K \\
0   & M_2 & m_Z c_{\beta} c_{W} & -m_Z s_{\beta} c_{W} & 0 &  0 \\
-m_Z c_{\beta} s_{W} & m_Z c_{\beta} c_{W} & 0 & -\mu_{eff} & -\mu_{\lambda} s_{\beta}& Q_{H_d}^{\prime} m_{v} c_{\beta}\\
m_Z s_{\beta} s_{W} & -m_Z s_{\beta} c_{W} & -\mu_{eff} & 0 & -\mu_{\lambda} c_{\beta} & Q_{H_u}^{\prime} m_{v} s_{\beta}\\
 0 & 0 & -\mu_{\lambda} s_{\beta} & -\mu_{\lambda} c_{\beta} & 0 & Q_S^{\prime}m_s\\
M_K & 0 & Q_{H_d}^{\prime} m_{v} c_{\beta} & Q_{H_u}^{\prime} m_{v} s_{\beta} & Q_S^{\prime}m_s & M_1^{\prime}\\
\end{array}\right) \nonumber \\
\eea
with gaugino mass parameters $M_1$ , $M_2$ , $M_1^{\prime}$ and
 $M_K$ \cite{Choi:2006fz} for $\tilde B$ ,  $\tilde W^3$ ,  ${\tilde B}^{\prime}$ and
  $\tilde B - {\tilde B}^{\prime}$ mixing respectively. There
  arise two additional mixing parameters after electroweak
  breaking:
\bea m_{v} = g_{Y^{\prime}} v ~~~~~ and ~~~~~ m_s = g_{Y^{\prime}}
v_s \eea Moreover, supersymmetric higgsino mass and
doublet-singlet higgsino mixing masses are generated to be \bea
\mu_{eff} = h_S \frac{v_S}{\sqrt{2}} ~~~~~~~,~~~~~~~ \mu_{\lambda}
= h_S \frac{v}{\sqrt{2}} \eea where $v=\sqrt{v_u^2 + v_d^2}$. The
neutralino mass matrix can be diagonalized by a unitary matrix
such that \bea N^{\dagger} M_{\chi^0} N= diag({\tilde
m}_{\chi_1^0} ,..., {\tilde m}_{\chi_6^0} ) \eea The additional
neutralino mass eigenstates due to new higgsino and gaugino
 fields encode effects of $U(1)^{\prime}$ models wherever neutralinos play a
 role such as magnetic and electric dipole moments.

In fact, the neutralino-sfermion exchanges contribute to EDMs of
quarks and leptons as follows:  \bea
\frac{d_{f-\chi^0}^E}{e}=\frac{\alpha_{EM}}{4\pi
\sin^2{\theta_W}}\sum_{k=1}^{2}\sum_{i=1}^{6}\,Im(\eta_{fik})\frac{\tilde{m}_{\chi_i^0}}{m_{\tilde{f}_k}^2}Q_{\tilde{f}}
B\left(\frac{\tilde{m}_{\chi_i^0}^2}{m_{\tilde{f}_k}^2}\right )
\eea where the neutralino vertex is, \bea
\eta_{fik}&=&\left[-\sqrt{2}\{\tw(Q_f-T_{3f})N_{1i}+\frac{g_{Y^{\prime}}}{g_2}Q_{f_L}^{\prime}N_{6i}
+T_{3f}N_{2i}\}D^{\star}_{f1k}-\kappa_f\,N_{bi}D^{\star}_{f2k}\right]\nonumber\\
&\times&(\sqrt{2}(\tw
\,Q_f\,N_{1i}+\frac{g_{Y^{\prime}}}{g_2}Q_{f_R}^{\prime}N_{6i})D_{f2k}-\kappa_f
N_{bi}D_{f1k})
\eea
and
\bea
\kappa_u = \frac{m_u}{\sqrt{2} M_W \sin{\beta}}~~~~,~~~~
\kappa_{d,e} = \frac{m_{d,e}}{\sqrt{2} M_W \cos{\beta}}
\eea
\bea
A(x)=\frac{1}{2(1-x)^2}\left(3-x+\frac{2\,\ln{x}}{1-x}\right)~,~
B(x)=\frac{1}{2(x-1)^2}\left(1+x+\frac{2x\ln{x}}{1-x}\right)
\eea
Since $H_u$ and $H_d$ couple fermions differently due to their hypercharges,
the $b$ index in neutralino diagonalizing matrix must be carefully chosen in numerical analysis.

\subsection{Chargino sector}
Unlike the Higgs and Neutralino sectors, chargino sector is
structurally unchanged in $U(1)^{\prime}$ models compared to MSSM.
However, chargino mass eigenstates become dependent upon
$U(1)^{\prime}$ breaking scale through $\mu_{eff}$ parameter in
their mass matrix: 
\bea 
M_{\chi^\pm}= \left(\begin{array}{c c}
M_2 & M_W \sqrt{2} \sin{\beta} \\
M_W \sqrt{2} \cos{\beta} & \mu_{eff} \\
\end{array} \right)
\eea 
which can be diagonalized by biunitary transformation
\bea U^{\star} M_{\chi^\pm} V^{-1} = diag(\tilde{m}_{\chi_1^+} ,
\tilde{m}_{\chi_2^+}) 
\eea 
where $U$ and $V$ are unitary mixing
matrices. Since the chargino sector is structurally the same as
with the MSSM, the fermion EDMs through fermion-sfermion-chargino
interactions are given by 
\bea
\frac{d_{e-\chi^\pm}^E}{e}=\frac{\alpha_{EM}}{4\pi
\sin^2{\theta_W}}\frac{\kappa_e}{m_{\tilde{\nu_e}}^2}\sum_{i=1}^2\,\tilde{m}_{\chi_i^+}
Im(U_{i2}^{\star}\,V_{i1}^{\star})A\left(\frac{\tilde{m}_{\chi_i^+}^2}{{m_{\tilde{\nu_e}^2}}}\right)
\eea

\bea
\frac{d_{d-\chi\pm}^E}{e}=-\frac{\alpha_{EM}}{4\pi
\sin^2{\theta_W}}\sum_{k=1}^{2}\sum_{i=1}^{2}Im(\Gamma_{dik})\frac{\tilde{m}_{\chi_i^+}}{m_{\tilde{u}_k}^2}
\left[Q_{\tilde{u}}B\left(\frac{\tilde{m}^2_{\chi_i^+}}{m_{\tilde{u}_k}^2}\right)+
(Q_d-Q_{\tilde{u}})A\left(\frac{\tilde{m}^2_{\chi_i^+}}{m_{\tilde{u}_k}^2}\right)\right]
\eea

\bea
\frac{d_{u-\chi^\pm}^E}{e}=-\frac{\alpha_{EM}}{4\pi
\sin^2{\theta_W}}\sum_{k=1}^{2}\sum_{i=1}^{2}Im(\Gamma_{uik})\frac{\tilde{m}_{\chi_i^+}}{m_{\tilde{d}_k}^2}
\left[Q_{\tilde{d}}B\left(\frac{\tilde{m}^2_{\chi_i^+}}{m_{\tilde{d}_k}^2}\right)+
(Q_u-Q_{\tilde{d}})A\left(\frac{\tilde{m}^2_{\chi_i^+}}{m_{\tilde{d}_k}^2}\right)\right]
\eea
where the chargino vertices are,
\bea
\Gamma_{uik} = \kappa_u V_{i2}^{\star} D_{d1k} (U_{i1}^{\star} D_{d1k}^{\star} - \kappa_d U_{i2}^{\star} D_{d2k}^{\star})
\eea
\bea
\Gamma_{dik} = \kappa_d U_{i2}^{\star} D_{u1k} (V_{i1}^{\star} D_{u1k}^{\star} - \kappa_u V_{i2}^{\star} D_{u2k}^{\star})
\eea

\subsection{Electron and Neutron EDMs}
Total EDMs for electron and neutron is therefore the sum of all
individual interactions, the electron EDM arises from CP-violating
1-loop diagrams with the neutralino and chargino exchanges \bea
d^E_e = d^E_{e-\chi^0} + d^E_{e-\chi^{\pm}} \eea While studying
neutron EDMs, besides neutralino and chargino diagrams, 1-loop
gluino
 exchange contribution must also be taken into account, thus the EDM for quark-
 squark-gluino interaction can be written as;
 \bea
 \frac{d^E_{q-\tilde{g}}}{e}=-\frac{2\alpha_s}{3\pi}\sum_{k=1}^{2}Im(\Gamma_{q}^{1k})\frac{m_{\tilde{g}}}{m_{\tilde{q}_k}^2} Q_{\tilde{q}} B\left(\frac{m_{\tilde{g}}^2}{m_{\tilde{q}_k}^2}\right)
 \eea
 with the gluino vertex,
 \bea
 \Gamma_{q}^{1k}=D_{q2k} D^{\star}_{q1k}
 \eea
However, for neutron EDM there are additionally two other
contributions arising from quark chromoelectric dipole moment of
quarks;
\bea{d}^C_{q-\tilde{g}}=\frac{g_s \alpha_s}{4\pi}\sum^2_{k=1}Im(\Gamma_{q}^{1k})\frac{m_{\tilde{g}}}{m_{\tilde{q}_k}^2}C\left(\frac{m_{\tilde{g}}^2}{m_{\tilde{q}_k}^2}\right)
\eea

\bea {d}^C_{q-\chi^0}=\frac{g_s g^2}{16 \pi^2}\sum^{2}_{k=1}
\sum^{6}_{i=1}
Im(\eta_{qik})\frac{\tilde{m}_{\chi^0_i}}{m^2_{\tilde{q}_k}} B
\left( \frac{\tilde{m}^2_{\chi^0_i}}{m^2_{\tilde{q}_k}}\right) \eea

\bea {d}^C_{q-\chi^{\pm}}=\frac{- g_s g^2}{16 \pi^2}\sum^{2}_{k=1}
\sum^{2}_{i=1}
Im(\Gamma_{qik})\frac{\tilde{m}_{\chi^{\pm}_i}}{m^2_{\tilde{q}_k}} B
\left( \frac{\tilde{m}^2_{\chi^{\pm}_i}}{m^2_{\tilde{q}_k}}\right)
\eea
where,
\bea
C(x)=\frac{1}{6(x-1)^2}\left(10x-26+\frac{2x\ln{x}}{1-x}-\frac{18\ln{x}}{1-x}\right)
\eea
and the CP violating dimension-six operator from 2-loop gluino-top-stop diagram is 
\bea
d^G=-3\alpha_s m_t \left(\frac{g_s}{4\pi}\right)^3Im(\Gamma_t^{12})\frac{z_1-z_2}{m_{\tilde{g}}^3}H(z_1,z_2,z_t)
\eea
with
\bea
z_i=\left(\frac{M_{\tilde{t}_i}}{m_{\tilde{g}}}\right)^2 ~~~~,~~~~ z_t=\left(\frac{m_t}{m_{\tilde{g}}}\right)^2
\eea
and the 2-loop function is given by \cite{Dai:1990xh}
\bea
H(z_1,z_2,z_t) = \frac{1}{2}\int_0^1 dx \int_0^1 du \int_0^1 dy\,
x\,(1-x)\,u\, \frac{N_1 N_2}{D^4} \eea 
with 
\bea
N_1&=&u\,(1-x) + z_t\,x\,(1-x)(1-u)-2u\,x\,[z_1\,y+z_2(1-y)] , \nonumber \\
N_2&=&(1-x)^2(1-u)^2+u^2-\frac{1}{9}x^2(1-u)^2 , \nonumber \\
D&=&u\,(1-x)+z_t\,x\,(1-x)(1-u)+u\,x\,[z_1\,y+z_2\,(1-y)]
\eea
Therefore total neutron EDM is written with the help of
non-relativistic
 $SU(6)$ coefficients of chiral quark model \cite{Abel:2001vy}
\bea d_n = \frac{1}{3}(4\,d_d-d_u) \eea in which all the
contributions are gathered into $u$ and $d$ quark interactions
\bea
d^E_u = d^E_{u-\chi^0} + d^E_{u-\chi^\pm} + d^E_{u-\tilde g} + d^C_{u-\chi^0} + d^C_{u-\chi^\pm} + d^C_{u-\tilde g} + d^G
\eea
\bea
d^E_d = d^E_{d-\chi^0} + d^E_{d-\chi^\pm} + d^E_{d-\tilde g} + d^C_{d-\chi^0} + d^C_{d-\chi^\pm} + d^C_{d-\tilde g} + d^G
\eea
The above analysis is at the electroweak scale
and the evolution of $d^{E,C,G}$'s down to hadronic scale is
accomplished via Naiv\"{e} Dimensional Analysis 
\bea
d_q = \eta^E d^E_q + \eta^C \frac{e}{4 \pi} d^C_q + \eta^G \frac{e \Lambda}{4 \pi} d^G
\eea
where  the QCD correction
factors are $\eta^E=1.53$, $\eta^C \simeq  3.4$ and $\Lambda \simeq
1.19$ GeV is the chiral symmetry breaking scale \cite{Ibrahim:1997gj}.

For the sake of generality, we give all the formulae which may contribute to electron and neutron EDM's,
however, depending on the origin of CP violating phases, some of above equations may yield no contributions to the EDM's, as in our numerical analysis we considered only one CP-odd phase corresponding to complex bino (and bino-prime) mass, for simplicity. Therefore in our analysis
contributions of gluinos for quark-squark-gluino interaction ($d_{q-\tilde g}^E$), chromoelectric dipole moment of quarks ($d^C_{q-\tilde{g}}$) and the CP violating dimension-six operator from the 2-loop gluino-top-stop diagram ($d^G$) will be missing. Care should be paid to the point that this phase can only provide a subleading contribution to the neutron EDM, for a complete treatment those missing contributions should be added too.

\subsection{Numerical Analysis}

In this part we will perform a detailed numerical study of various
$E(6)$--based $U(1)^{\prime}$ models in regard to their
predictions for electron and neutron EDMs. We will compare the
models given in Tab. \ref{tab1} with each other and with the MSSM.
In doing this, we consider bino (and bino-prime) mass to be complex and assume the
rest of the parameters as real quantities (though this
simplification might seem somewhat unrealistic we expect that
results can still reveal certain salient features in such models).

During the  analysis, to respect the collider bounds, we require the
masses satisfy
\begin{equation}
m_h>90 ,\, m_{sfermions}>100,\,m_{\chi^\pm_1}>105,\,M_{Z^\prime}>700
\end{equation}
(all in GeV) and the $Z-Z^\prime$ mixing angle to be less than
$3\,\times\,10^{-3}$. Bounds from naturalness and perturbativity
constraint are respected by considering $0.1\leq\,h_s\,\leq0.75$
\cite{Masip:1999mk,Suematsu:1997tv,Demir:2003ke}. Additionally, to make $Z^\prime$ sufficiently  heavy
$v_s$ is scanned up to 10 TeV and low $\tan\beta$ regime  is
analyzed which is the preferred  domain for the  models and for
which consideration of stop corrections suffice.

\begin{figure}[ht!]
\begin{center}
\includegraphics[scale=1,height=15cm]{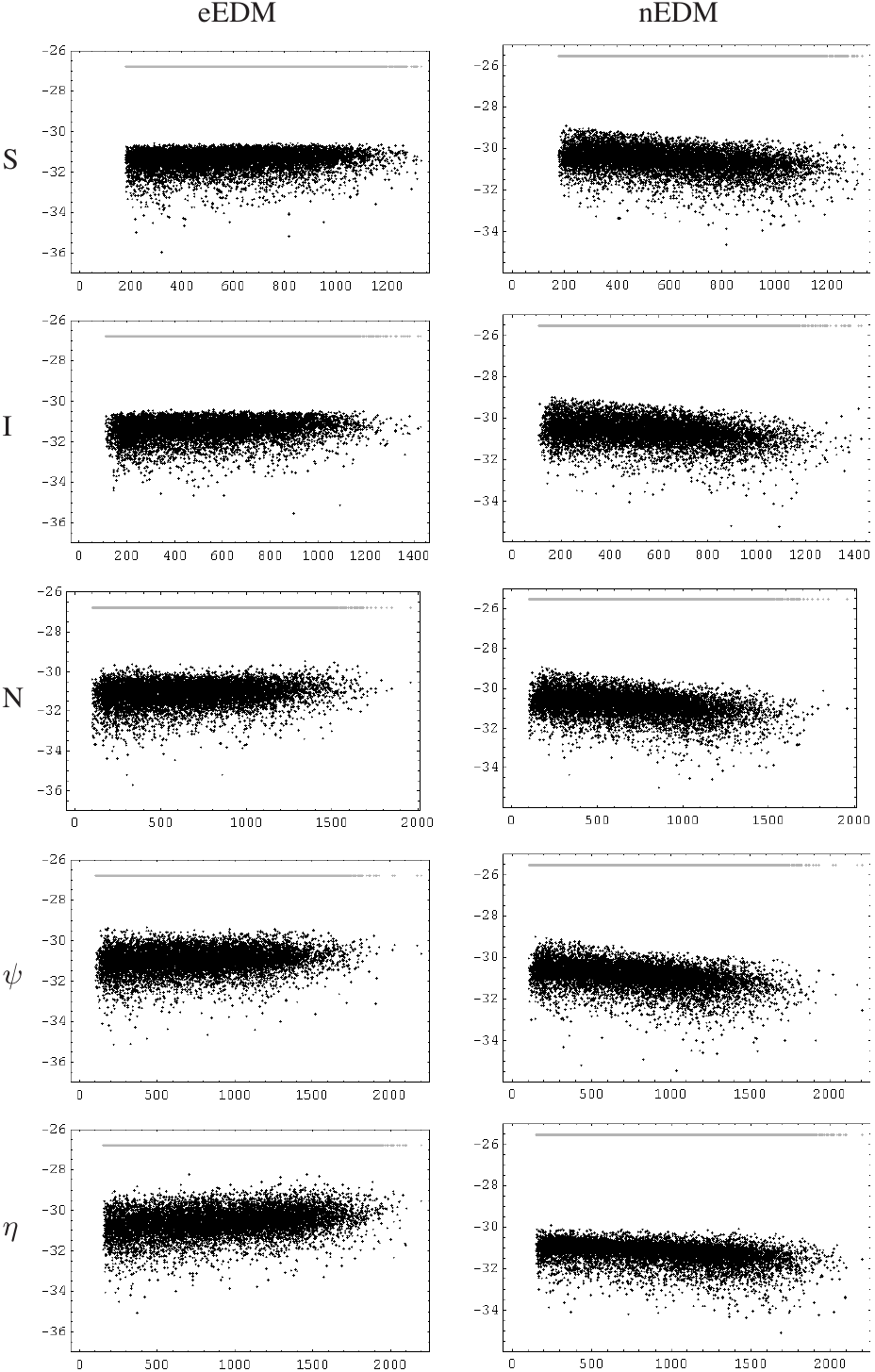}
\end{center}
\caption{ $\mu_{eff}$ versus eEDM (left panels) and nEDM (right
panels) in $U(1)^\prime$ models (top to bottom: S, I, N, $\psi$ and
$\eta$ models). As inputs, all trilinears are scanned in -2 to 2
TeV, all sfermions are scanned in 0.5 to 1 TeV separately. The
resulting data sets are used to obtain in every model with
$\tan\beta=3$. Absolute value of  EDM predictions are given in
$\log_{10}$ base, $\mu_{eff}$ values are given in GeVs. Straight
lines in this and following figures denote corresponding eEDM and nEDM
experimental constraints \cite{Regan:2002ta,Baker:2006ts}.} \label{fig2}
\end{figure}

Imprints of different $U(1)^\prime$ models related with electron
and neutron EDM reactions  are presented in Fig. \ref{fig2}. This
figure depicts variations of EDMs with $\mu_{eff}$ in S, I, N ,
$\psi$ and $\eta$ models. In this figure and in the followings,
since we did not take into consideration renormalization group
running, we scanned the related parameters randomly. But we
carefully used the same  data points in each of the models. As can
be seen from Fig. \ref{fig2}, with increasing $\mu_{eff}$, eEDM
(left panels) predictions start to raise from S to $\eta$ model.
Additionally, as the effective $\mu$ parameter deviates from the
EW scale, eEDM predictions seem promising to bound the effective
$\mu$ term in $\eta$ and $\psi$ models. But when it comes to nEDM
(right panels) as  the $\mu_{eff}$ increases predictions for
neutron EDM decreases from S to $\eta$ model, respectively. In
other words, in terms of the difference between electron and
neutron EDM predictions, the $\eta$ model is the most striking one
and the S model is the mildest model.

\begin{figure}[ht!]
\begin{center}
\includegraphics[scale=1,height=15cm]{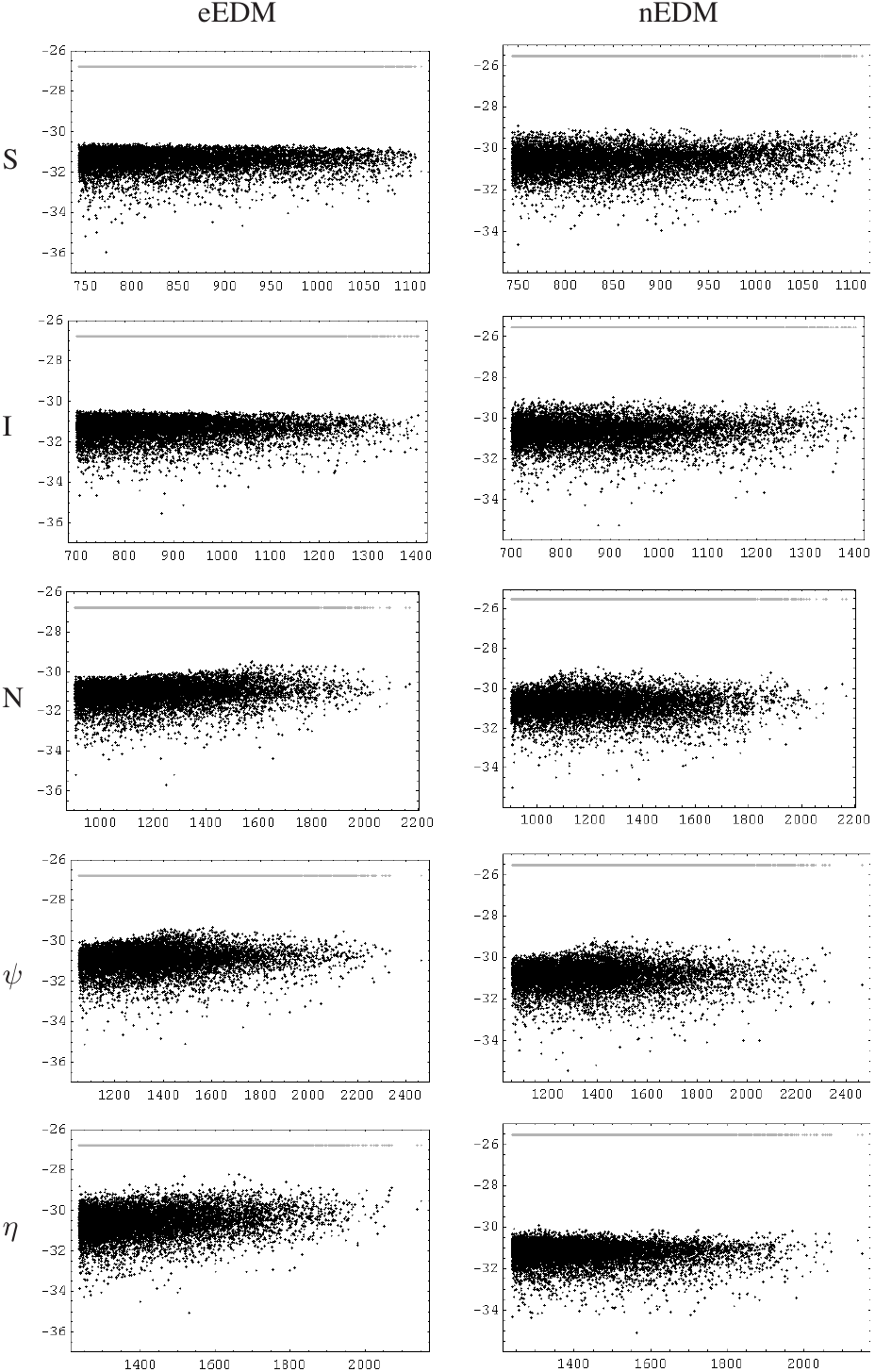}
\end{center}
\caption{ $M_{Z^\prime}$ versus eEDM (left panels) and nEDM (right
panels) in $U(1)^\prime$ models, as in Fig. \ref{fig2}.}
\label{fig3}
\end{figure}

It is also useful to probe how  EDM predictions vary with the mass
of $Z^\prime$ boson, which is given in Fig. \ref{fig3}. The left
$\eta$ panel of Fig \ref{fig3} shows that it may be possible to
bound $Z^\prime$ mass from above once the eEDM predictions  near
the present experimental value (at least for certain range of
parameters), whereas some models like S and I do not seem to react
significantly to this variation. The most sensitive models to
bound $Z^\prime$ mass using the eEDM results are $\eta$, $\psi$
and N models.  On the other hand, it may also be possible to bound
the mass of $Z^\prime$ in S model using the nEDM measurements, as
can be seen from the bottom S panel of Fig. \ref{fig3}.

Our next figure is Fig. \ref{fig4} in which electron and neutron EDM
predictions are presented for the MSSM and for the aforementioned
$U(1)^\prime$ models against variations in the phase of bino. In S
and I models eEDM predictions are generally well below the MSSM
predictions. On the other hand, in $\eta$ model it is possible to
get lower predictions for nEDM. Notice that while majority  of the
points obtained are above the MSSM predictions there are regions
where it is possible to obtain smaller EDM values for both of the
electron and neutron (i.e. see the gray crosses  in N and $\psi$
panels).

\begin{figure}[ht!]
\begin{center}
\includegraphics[scale=1,height=15cm]{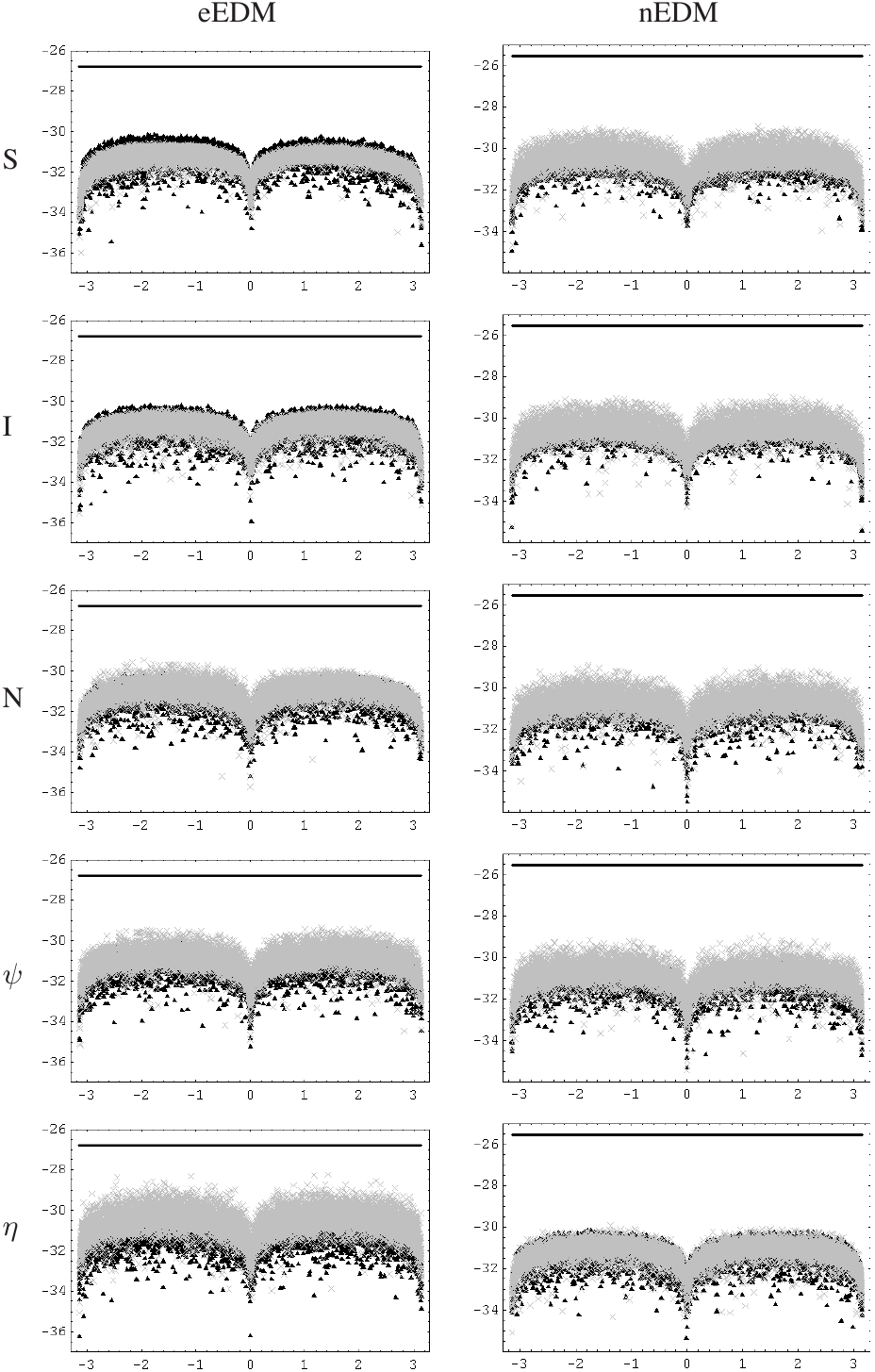}
\end{center}
\caption{The phase of  $M_{1}$ versus eEDM (left panels) and nEDM
(right panels) in $U(1)^\prime$ models. Here our shading convention
is such that dark triangles correspond to  MSSM and gray crosses are
for $U(1)^\prime$ models. Inputs are as in Fig. \ref{fig2}.}
\label{fig4}
\end{figure}
\begin{figure}[ht!]
\begin{center}
\includegraphics[scale=1,height=20mm]{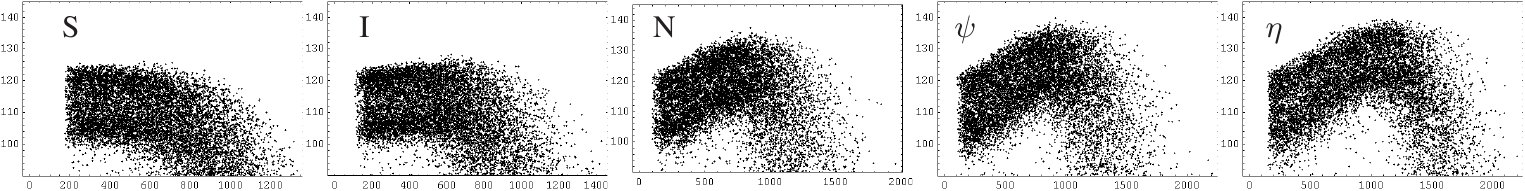}
\end{center}
\caption{Effective $\mu$ versus $m_{h}$  in $U(1)^\prime$ models
(All in GeVs). Inputs are the same with Fig. \ref{fig2}.}
\label{fig5}
\end{figure}

As can be deduced from the previous figures there is a hierarchy
among the models. This situation is also shared by  the mass of
the lightest Higgs boson. We provide Fig. \ref{fig5} in which mass
of the lightest Higgs boson is plotted against variations of
$\mu_{eff}$. Here again, predictions for the mass of the lightest
Higgs boson are in an order increasing from S to $\eta$ model.
Notice that while the LEP2 bound on SM like Higgs boson confines
its mass to be larger than 114 GeV it can not be used directly in
$U(1)^\prime$ models, so we accepted 90 GeV as the lower bound. But
all of the models are capable of satisfying $m_h>114$ GeV.
Additionally, compared to the MSSM, in these $U(1)^\prime$ models
it  is possible to find larger $m_h$ predictions for $m_h$ i.e.
see $\eta$ or $\psi$ panels.

\begin{figure}[ht!]
\begin{center}
\includegraphics[scale=1,height=15cm]{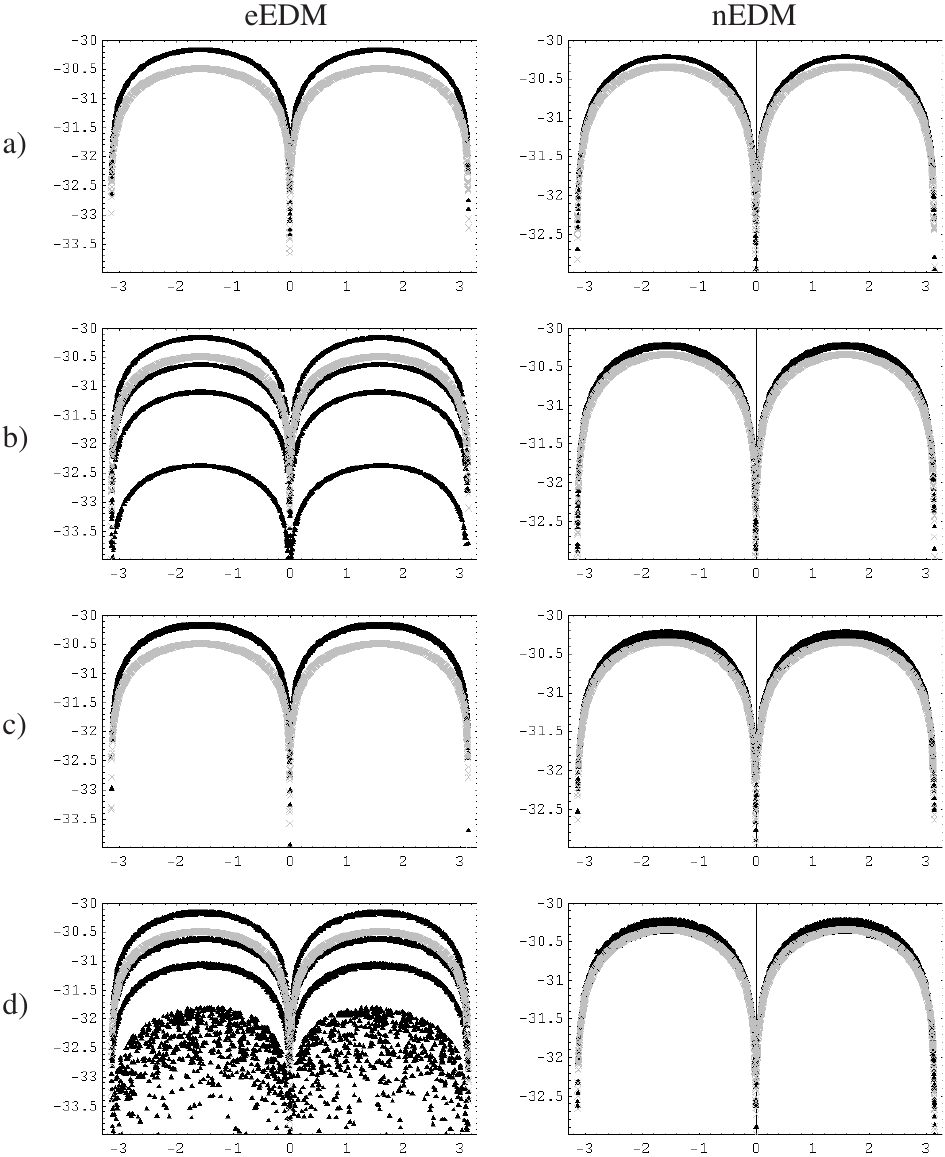}
\end{center}
\caption{The eEDM (left panels) and the nEDM (right panels) versus
argument of $M_1$ in N model (Dark triangles : MSSM, gray crosses :
N model). Here we fixed $\tan\beta=5$, $m_{\rm{sleptons}}=400$ GeV,
$m_{\rm{squarks}}=750$ GeV, all trilinars=-1500 GeV, $M_2=190$ GeV
($M_1=0.56\,M_2$, $M_3=2.8\,M_2$) In panel a) mixing angle $\chi=0$,
$M_{YX}=0$, b) mixing angle $\chi=-0.3,-0.2,-0.1,0$ and $M_{YX}=0$,
c) mixing angle $\chi=0$ but $M_{YX}$ scanned randomly in 0 to 0.5
TeV d) $\chi=-0.3,-0.2,-0.1,0$ and $M_{YX}$ scanned randomly in 0 to
0.5 TeV. Notice that $M_{YX}\sim\,M_K$ for small $\chi$ values as in
our cases (see \cite{Choi:2006fz} for details)} \label{fig6}
\end{figure}

Another important issue worth noticing within these models is the
possibility of kinetic mixing. As should be predicted it modifies
EDM predictions (as well as many other properties of the models) in
accordance with its magnitude. To give a concrete example of its
impact, we selected N model for which  eEDM and nEDM predictions are
generally  larger than the MSSM. So, we provide Fig. \ref{fig6} for
electron and neutron EDMs. As can be seen  the very  figure, even
very small values of the kinetic mixing angle (i.e. $\chi$=-0.1) can
yield sizable variations for the EDM predictions of the electron,
but, its impact on the neutron EDM is rather small. Meanwhile,
nonzero choices of the mass terms $M_K$ (see the $c$ panels) can
also reduce both of the eEDM and nEDM predictions. When both of the
$\chi$ and $M_K$ are in charge (see the $d$ panels), we see that,
both of the eEDM and nEDM predictions in the N model can be smaller
than the MSSM predictions.

A rather interesting effect of the kinetic mixing can be
investigated on the composition of the LSP candidate of the
$U(1)^\prime$ models. For the selected range of the parameters,
all $U(1)^\prime$ models share the same LSP candidate with the
MSSM, which is bino. But also notice that singlino dominated
neutralino can be a good candidate for the LSP
\cite{Nakamura:2006ht,Suematsu:2005bc}, for this kind of models.

\begin{figure}[ht!]
\begin{center}
\includegraphics[scale=1,height=3.5cm]{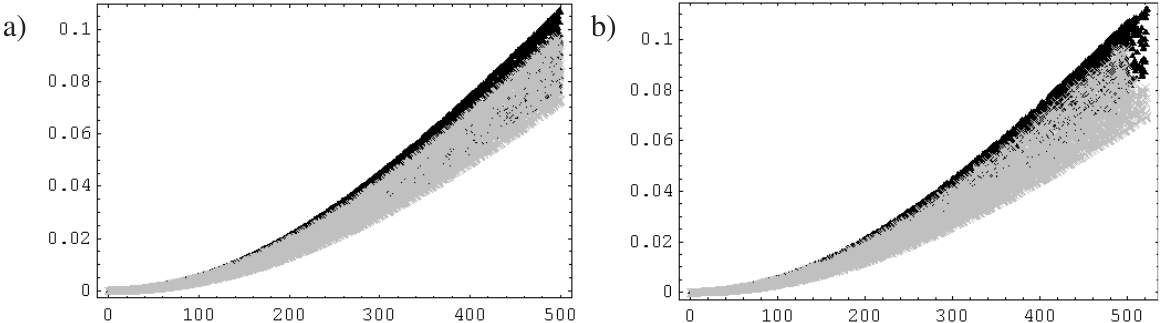}
\end{center}
\caption{Singlino (gray crosses: $\rm{|N_{1,5}|^2}$) and
$Z^\prime$-ino (dark tringles: $\rm{|N_{1,6}|^2}$) compositions of
the lightest neutralino against $M_K$ in N model. Inputs are from
c) and d) panels of Fig. \ref{fig6}. (for a) and b) panels they are
of the order $10^{-7}$).} \label{fig7}
\end{figure}

\begin{figure}[ht!]
\begin{center}
\includegraphics[scale=1,height=3.5cm]{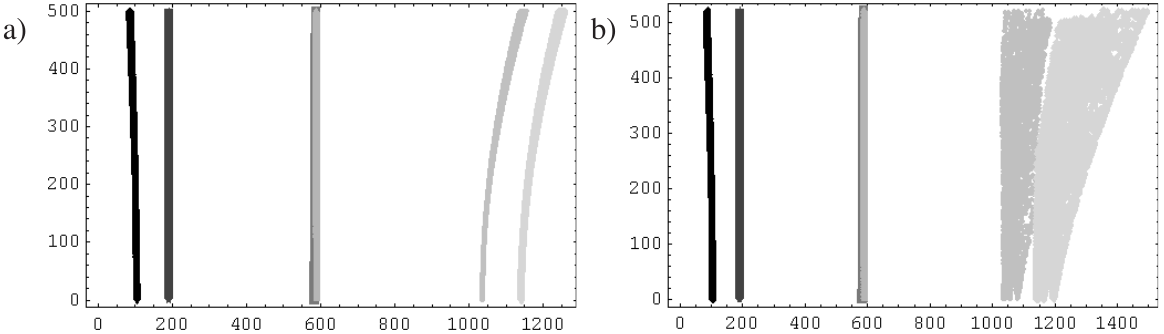}
\end{center}
\caption{Neutralino masses versus $M_K$ corresponding to the same
panels of Fig. \ref{fig7} (All in GeVs).} \label{fig8}
\end{figure}

In our domain, without the kinetic mixing its composition can be
expected to be very similar to the MSSM's lightest neutralino. This
can be inferred from  Fig. \ref{fig7}  where singlino (gray crosses)
and $Z^\prime$-ino (dark triangles) compositions of the LSP
candidate are plotted against varying $M_K$ with (left panel) and
without (right panel) the kinetic mixing scanned randomly in
[-0.3,0]. Notice that when $M_K\sim0$ GeV, even if the kinetic
mixing is turned on, the composition of the LSP candidate can not be
expected to be very different from the MSSM.
For a clear picture of this phenomena we support Figs.  \ref{fig6} and
\ref{fig7} with  Fig. \ref{fig8}, where the mass eigenvalues of the
N model neutralinos are plotted against varying  $M_K$ with (panel
b)) and without (panel a)) mixing angle. As can be seen from Fig
\ref{fig8}, mass of the LSP candidate of the related model is
sensitive to $M_K$. This tendency reduces as we go away from the
lightest neutralino up to 5th and 6th neutralinos. For those two
heavy neutralinos impact of nonzero mixing angle can dominate the
effect of $M_K$ if both of them are in charge (see panel b) of Fig
\ref{fig8}). For the selected range of parameters lightest
neutralino is very similar to the MSSM's neutralino as far as the
mentioned variables are off; when they are on, their corresponding
impact on the composition and on the mass of the lightest neutralino
can be $\sim$ 10-20 $\%$ as can be seen from the very figures.

\begin{figure}[ht!]
\begin{center}
\includegraphics[scale=1,height=15cm]{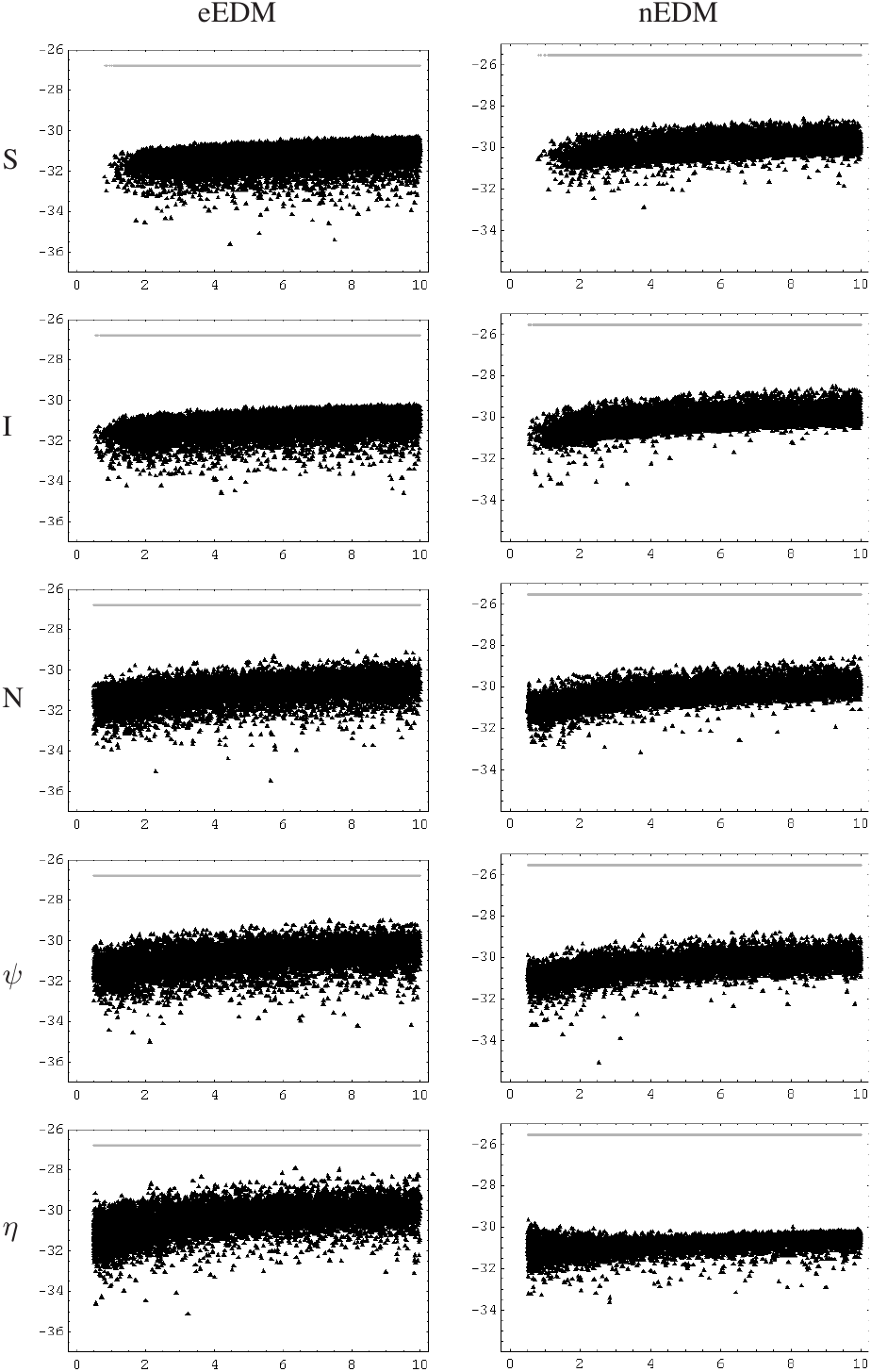}
\end{center}
\caption{$\tan\beta$ versus eEDM (left panels) and nEDM (right
panels) predictions in different $U(1)^\prime$ models. We used the
conventions of Fig. \ref{fig3}. Here again straight lines denotes
the corresponding EDM bounds.} \label{fig9}
\end{figure}

Our last figure is Fig. \ref{fig9} where  we present $\tan\beta$
dependencies of the electron and neutron EDMs. Here $\tan\beta$ is
scanned up to 10 and the most striking difference between the MSSM
and $U(1)^\prime$ models, for the models under concern, turns out
to be the smallness of  $\tan\beta$ (can be as small as 0.5),
which is ruled out for the MSSM. Additionally, for most of the
models eEDM and nEDM predictions decrease with decreasing
$\tan\beta$ as in the MSSM. The only exception to this observation
is found for $\eta$ model where the sensitivity of eEDM
predictions are very small. But, in general, this common  tendency
of $U(1)^\prime$ models show that it is easier to evade EDM
constraints in such models where $\tan\beta\sim1$ is actually the
natural value.

As can be seen from the figures presented in this section, we did not
try to constrain complex phases but instead we tried to demonstrate 
the general tendencies in $U(1)^{\prime}$ models, and apparently all the
examples given here are well below the experimental bounds. 

\section{Conclusion}

In this work we have performed a study of EDMs (of electron and
neutron) in $U(1)^{\prime}$ models descending from $E(6)$ SUSY GUT.
With anticipated increase in precision of EDM measurements, our
results show that these models give rise to observable signatures
not shared by the MSSM. Indeed, $U(1)^{\prime}$ models generically
possess different predictions for EDMs compared to MSSM (see Fig.
4). This very feature provides a way of determining nature of the
supersymmetric model at the ${\rm TeV}$ scale via EDM measurements.

Apart from comparisons with the MSSM,  different $E(6)$--based
$U(1)^{\prime}$ models are found to have different predictions for
various observables studied in the text. Indeed, sensitivity of
EDMs to $\mu$ parameter (see Fig. 2), to $Z^{\prime}$ mass (see
Fig. 3), and to $\tan\beta$ are different for different models.
Furthermore, eEDM and nEDM are found to exhibit different
dependencies in each case. These features establish the fact that,
once precise measurements are attained (presumably at a
high-energy linear collider) one can determine likely breaking
directions for $E(6)$ grand unified group down to that of the
MSSM.

Fig. 6  makes it clear that the soft-breaking mass that mix $U(1)_Y$
and $U(1)^{\prime}$ gauginos is a sensitive source of EDMs. Indeed,
as happens in models of paraphotons, entire matter can be neutral
under $U(1)^{\prime}$ symmetry yet such a kinetic mixing (that mix
gauge bosons and gauginos ) can exist and can have important
implications. These figures make it clear that EDMs vary
significantly with this parameter.

Also interesting are the predictions of different $U(1)^{\prime}$
models for $m_h$ (which is plotted against $\mu_{eff}$ in Fig. 5).
Indeed, both range and shape of the allowed domain are different
for different models, and this feature also helps determining the
correct model (of $E(6)$ origin) once precise measurements of
associated quantities are available.

It is not surprising that these models can have important
implications also for FCNC observables (including their CP
asymmetries) \cite{langackerx}. Moreover, the EDMs discussed above
can be correlated with the CP asymmetries (of $B$ meson decays
\cite{correlate}) or with the Higgs sector itself
\cite{correlate2} so as to further bound such models with the
information available from $B$ factories and Tevatron. This kind
of analysis will be given elsewhere.

To conclude, the problem of CP violation (in particular EDMs) is a
particularly important issue of $U(1)^{\prime}$ models for various
reasons, most notably, the approximate reality of the effective
$\mu$ parameter. Analyses of various observables (including the
FCNC ones) can shed further light on the origin and structure of
such models.

\section{Acknowledgments}
We all would like to thank to D. A. DEM{\.{I}}R for his
contributions with inspiring and illuminating discussions in various
stages of this work.


\begin{thebibliography}{99}
\bibitem{muprob}
%\bibitem{Kim:1983dt}
J.~E.~Kim and H.~P.~Nilles,
%``The Mu Problem And The Strong CP Problem,''
Phys.\ Lett.\ B {\bf 138}, 150 (1984);
%%CITATION = PHLTA,B138,150;%%
D.~Suematsu and Y.~Yamagishi,
%``Radiative symmetry breaking in a supersymmetric model with an extra U(1),''
Int.\ J.\ Mod.\ Phys.\ A {\bf 10}, 4521 (1995)
[arXiv:hep-ph/9411239];
%%CITATION = HEP-PH 9411239;%%
M.~Cvetic and P.~Langacker,
%``New Gauge Bosons from String Models,''
Mod.\ Phys.\ Lett.\ A {\bf 11}, 1247 (1996)
[arXiv:hep-ph/9602424];
%%CITATION = HEP-PH 9602424;%%
V.~Jain and R.~Shrock,
%``U(1)-A models of fermion masses without a mu problem,''
arXiv:hep-ph/9507238;
%%CITATION = HEP-PH 9507238;%%
Y.~Nir,
%``Gauge unification, Yukawa hierarchy and the mu problem,''
Phys.\ Lett.\ B {\bf 354}, 107 (1995) [arXiv:hep-ph/9504312].
%%CITATION = HEP-PH 9504312;%%

%\cite{Robinett:1981yz}
\bibitem{Robinett:1981yz}
  R.~W.~Robinett and J.~L.~Rosner,
  %``Prospects For A Second Neutral Vector Boson At Low Mass In SO(10),''
  Phys.\ Rev.\  D {\bf 25}, 3036 (1982)
  [Erratum-ibid.\  D {\bf 27}, 679 (1983)].
  %%CITATION = PHRVA,D25,3036;%%

%\cite{Robinett:1982tq}
\bibitem{Robinett:1982tq}
  R.~W.~Robinett and J.~L.~Rosner,
  %``Mass Scales In Grand Unified Theories,''
  Phys.\ Rev.\  D {\bf 26}, 2396 (1982).

%\cite{Langacker:1984dc}
\bibitem{Langacker:1984dc}
  P.~Langacker, R.~W.~Robinett and J.~L.~Rosner,
  %``New Heavy Gauge Bosons In P P And P Anti-P Collisions,''
  Phys.\ Rev.\  D {\bf 30}, 1470 (1984).
  %%CITATION = PHRVA,D30,1470;%%

%\cite{Cvetic:1995rj}
\bibitem{Cvetic:1995rj}
  M.~Cvetic and P.~Langacker,
  %``Implications of Abelian Extended Gauge Structures From String Models,''
  Phys.\ Rev.\  D {\bf 54}, 3570 (1996)
  [arXiv:hep-ph/9511378].
  %%CITATION = PHRVA,D54,3570;%%

%\cite{Cvetic:1996mf}
\bibitem{Cvetic:1996mf}
  M.~Cvetic and P.~Langacker,
  %``New Gauge Bosons from String Models,''
  Mod.\ Phys.\ Lett.\  A {\bf 11}, 1247 (1996)
  [arXiv:hep-ph/9602424].
  %%CITATION = MPLAE,A11,1247;%%

\bibitem{Demir:2005kg}
  D.~A.~Demir, L.~Solmaz and S.~Solmaz,
  %``LEP indications for two light Higgs bosons and U(1)' model,''
  Phys.\ Rev.\  D {\bf 73}, 016001 (2006)
  [arXiv:hep-ph/0512134].

\bibitem{Suematsu:1998wm}
  D.~Suematsu,
  %``Vacuum structure of the mu-problem solvable extra U(1) models,''
  Phys.\ Rev.\  D {\bf 59} (1999) 055017
  [arXiv:hep-ph/9808409].

\bibitem{King:2005jy}
  S.~F.~King, S.~Moretti and R.~Nevzorov,
  %``Theory and phenomenology of an exceptional supersymmetric standard
  %model,''
  Phys.\ Rev.\  D {\bf 73} (2006) 035009
  [arXiv:hep-ph/0510419].

\bibitem{Demir:2005ti}
  D.~A.~Demir, G.~L.~Kane and T.~T.~Wang,
  %``The minimal U(1)' extension of the MSSM,''
  Phys.\ Rev.\  D {\bf 72} (2005) 015012
  [arXiv:hep-ph/0503290].
  %%CITATION = PHRVA,D72,015012;%%

  \bibitem{Langacker:2008yv}
  P.~Langacker,
  %``The Physics of Heavy Z' Gauge Bosons,''
  arXiv:0801.1345 [hep-ph].

\bibitem{Choi:2006fz}
  S.~Y.~Choi, H.~E.~Haber, J.~Kalinowski and P.~M.~Zerwas,
  %``The neutralino sector in the U(1)-extended supersymmetric standard model,''
  Nucl.\ Phys.\  B {\bf 778} (2007) 85
  [arXiv:hep-ph/0612218].

%\cite{Demir:2007dt}
\bibitem{Demir:2007dt}
  D.~A.~Demir, L.~L.~Everett and P.~Langacker,
  %``Dirac Neutrino Masses from Generalized Supersymmetry Breaking,''
  Phys.\ Rev.\ Lett.\  {\bf 100}, 091804 (2008)
  [arXiv:0712.1341 [hep-ph]].
  %%CITATION = PRLTA,100,091804;%%
%\cite{AguilarSaavedra:2005pw}

%SPS1a'
\bibitem{AguilarSaavedra:2005pw}
  J.~A.~Aguilar-Saavedra {\it et al.},
  %``Supersymmetry parameter analysis: SPA convention and project,''
  Eur.\ Phys.\ J.\  C {\bf 46}, 43 (2006)
  [arXiv:hep-ph/0511344].
  %%CITATION = EPHJA,C46,43;%%

  %\cite{Demir:2003ke}
\bibitem{Demir:2003ke}
  D.~A.~Demir and L.~L.~Everett,
  %``CP violation in supersymmetric U(1)' models,''
  Phys.\ Rev.\  D {\bf 69}, 015008 (2004)
  [arXiv:hep-ph/0306240].
  %%CITATION = PHRVA,D69,015008;%%

\bibitem{Cvetic:1997ky}
M.~Cvetic, D.~A.~Demir, J.~R.~Espinosa, L.~L.~Everett and
P.~Langacker,
%``Electroweak breaking and the mu problem in supergravity models with an
%additional U(1),''
Phys.\ Rev.\ D {\bf 56}, 2861 (1997) [Erratum-ibid.\ D {\bf 58},
119905 (1998)] [hep-ph/9703317].

\bibitem{Dai:1990xh}
  J.~Dai, H.~Dykstra, R.~G.~Leigh, S.~Paban and D.~Dicus,
  %``CP VIOLATION FROM THREE GLUON OPERATORS IN THE SUPERSYMMETRIC STANDARD
  %MODEL,''
  Phys.\ Lett.\  B {\bf 237} (1990) 216
  [Erratum-ibid.\  B {\bf 242} (1990) 547].

\bibitem{Abel:2001vy}
  S.~Abel, S.~Khalil and O.~Lebedev,
  %``EDM constraints in supersymmetric theories,''
  Nucl.\ Phys.\  B {\bf 606} (2001) 151
  [arXiv:hep-ph/0103320].


\bibitem{Ibrahim:1997gj}
  T.~Ibrahim and P.~Nath,
  %``The neutron and the electron electric dipole moment in N = 1  supergravity
  %unification,''
  Phys.\ Rev.\  D {\bf 57} (1998) 478
  [Erratum-ibid.\  D {\bf 58} (1998\ ERRAT,D60,079903.1999\ ERRAT,D60,119901.1999) 019901]
  [arXiv:hep-ph/9708456].

  %\cite{Masip:1999mk}
\bibitem{Masip:1999mk}
  M.~Masip and A.~Pomarol,
  %``Effects of SM Kaluza-Klein excitations on electroweak observables,''
  Phys.\ Rev.\  D {\bf 60}, 096005 (1999)
  [arXiv:hep-ph/9902467].
  %%CITATION = PHRVA,D60,096005;%%

\bibitem{Suematsu:1997tv}
  D.~Suematsu,
  %``Effect on the electron EDM due to abelian gauginos in SUSY extra U(1)
  %models,''
  Mod.\ Phys.\ Lett.\  A {\bf 12} (1997) 1709
  [arXiv:hep-ph/9705412].

%\cite{Nakamura:2006ht}
\bibitem{Nakamura:2006ht}
  S.~Nakamura and D.~Suematsu,
  %``Supersymmetric extra U(1) models with a singlino dominated LSP,''
  Phys.\ Rev.\  D {\bf 75}, 055004 (2007)
  [arXiv:hep-ph/0609061].
  %%CITATION = PHRVA,D75,055004;%%

%\cite{Suematsu:2005bc}
\bibitem{Suematsu:2005bc}
  D.~Suematsu,
  %``Singlino dominating CDM in supersymmetric extra U(1) models,''
  Phys.\ Rev.\  D {\bf 73}, 035010 (2006)
  [arXiv:hep-ph/0511299].
  %%CITATION = PHRVA,D73,035010;%%

\bibitem{langackerx}
P.~Langacker and M.~Plumacher,
  %``Flavor changing effects in theories with a heavy Z' boson with family
  %non-universal couplings,''
  Phys.\ Rev.\  D {\bf 62}, 013006 (2000)
  [arXiv:hep-ph/0001204].
  %%CITATION = PHRVA,D62,013006;%%

  \bibitem{correlate}
  T.~M.~Aliev, D.~A.~Demir, E.~Iltan and N.~K.~Pak,
  %``The CP Asymmetry in $b \to s l~+ l~-$ Decay,''
  Phys.\ Rev.\  D {\bf 54}, 851 (1996)
  [arXiv:hep-ph/9511352].
  %%CITATION = PHRVA,D54,851;%%

\bibitem{correlate2}
D.~A.~Demir,
  %``Higgs boson couplings to quarks with supersymmetric CP and flavor
  %violations,''
  Phys.\ Lett.\  B {\bf 571}, 193 (2003)
  [arXiv:hep-ph/0303249];
  %%CITATION = PHLTA,B571,193;%%
A.~Dedes and A.~Pilaftsis,
  %``Resummed effective Lagrangian for Higgs-mediated FCNC interactions in the
  %CP-violating MSSM,''
  Phys.\ Rev.\  D {\bf 67}, 015012 (2003)
  [arXiv:hep-ph/0209306];
  %%CITATION = PHRVA,D67,015012;%%
M.~S.~Carena, A.~Menon, R.~Noriega-Papaqui, A.~Szynkman and
C.~E.~M.~Wagner,
  %``Constraints on B and Higgs physics in minimal low energy supersymmetric
  %models,''
  Phys.\ Rev.\  D {\bf 74}, 015009 (2006)
  [arXiv:hep-ph/0603106].
  %%CITATION = PHRVA,D74,015009;%%

\bibitem{Regan:2002ta}
  B.~C.~Regan, E.~D.~Commins, C.~J.~Schmidt and D.~DeMille,
  %``New limit on the electron electric dipole moment,''
  Phys.\ Rev.\ Lett.\  {\bf 88} (2002) 071805.
  %%CITATION = PRLTA,88,071805;%%

\bibitem{Baker:2006ts}
  C.~A.~Baker {\it et al.},
  %``An improved experimental limit on the electric dipole moment of the
  %neutron,''
  Phys.\ Rev.\ Lett.\  {\bf 97}, 131801 (2006)
  [arXiv:hep-ex/0602020].
  %%CITATION = PRLTA,97,131801;%%


\end{thebibliography}
\end{document}